\documentclass[prd,amsmath,notitlepage,twocolumn]{revtex4-1}
\usepackage{graphicx}
\usepackage{float}
\usepackage{epstopdf,cancel}
\usepackage{epsf,latexsym,bbm,euscript}
\usepackage{amssymb,amsmath}
\usepackage{mathtools} 
\usepackage{times,graphics}
\usepackage{soul,xcolor}

\def\6{{\langle}}
\def\9{{\rangle}}
\newcommand{\defeq}{\vcentcolon=}

\newcommand{\be}{\begin{equation}}
\newcommand{\ee}{\end{equation}}
\newcommand{\ba}{\begin{eqnarray}}
\newcommand{\ea}{\end{eqnarray}}

\newcommand{\mS}{{\mathrm{S}}}
 \newcommand{\mA}{{\mathrm{A}}}

\def\etal{\textit{et al.}}

\def\half{{\tfrac{1}{2}}}

\def\pad{{\partial}}

\def\sg{\textsl{g}}

\def\eC{\EuScript{C}}

\def\cO{\mathcal{O}}

\usepackage{url,hyperref}
\hypersetup{colorlinks,linkcolor={blue!55!black},citecolor={red!50!black},urlcolor={blue!45!black}}

\begin{document}

\title{Self-consistent description of a spherically-symmetric gravitational collapse}

\author{Daniel R. Terno}
\affiliation{Department of Physics \& Astronomy, Macquarie University, Sydney NSW 2109, Australia}

\begin{abstract}

In spherical symmetry, the total energy-momentum tensor near the apparent horizon is identified up to a single function of time from two assumptions:
a trapped region forms at  a finite time of a distant observer, and  values of two curvature scalars are finite at its boundary. In general relativity,  this energy-momentum tensor leads  to the
   unique limiting form of the metric.  The null energy condition is violated across the apparent horizon and
is satisfied in the vicinity of the inner apparent horizon.
As a result,   homogenous collapse models cannot describe the formation of a black hole.
         Properties of matter change discontinuously immediately after formation of a trapped region. Absolute values of  comoving density, pressure, and flux coincide at the apparent horizon.
          Thus,
         collapse of ideal fluids cannot lead to the formation of  black holes.    Moreover, these three quantities diverge at the expanding apparent horizon,  producing a regular (i.e., finite curvature)
          firewall.  This firewall is incompatible with quantum energy inequalities, implying that
trapped regions, once formed at some finite time of a distant observer, cannot grow. 

\end{abstract}
\maketitle

\section{Introduction}
Black holes  (BHs) are  envisaged as spacetime regions where  gravity is so strong that 
nothing, not even light, can   escape \cite{mit:783,curiel,visser:14}. 
 Mathematical black holes are solutions of the Einstein equations of general relativity (GR) \cite{he:book,fn:book,bambi}.
The salient property of these solutions is the event horizon that separates  the outside world from the  black hole interior.
  Astrophysical black hole (ABH) candidates are massive compact   dark objects.
  It is still not known how, when and  {if at all} they develop  the distinctive  features of the black holes of GR \cite{cp:na17,bh-map}.

Quantum effects and
uncertainty regarding the end result of the collapse \cite{bh-map,info,qua-mod}
 motivate investigations of exotic compact objects (ECOs) that  do not lead to formation of an event horizon and/or  singularity. Advances in instrumentation  make  studies
of spacetime close to ABHs possible \cite{eht-1:19}, focusing attention on the  observational differences between ECO  and
conventional black holes \cite{bh-map,qnm1,obs-t,cp:na17,cflv:18,fr:18}.

Event horizons are global teleological entities that are in principle unobservable
   \cite{visser:14,cp:na17}, and theoretical, numerical and observational studies focus on other characteristics of BHs \cite{faraoni:b,abh}.
A trapped region is a domain where both ingoing and outgoing  future-directed null
  geodesics emanating from a spacelike  two-dimensional surface  with spherical topology have negative expansion \cite{he:book,faraoni:b,abh,krishnam:14}. This local backward bending of light
prevents communications with the outside world.    The apparent horizon is the outer
boundary of the trapped region \cite{he:book, faraoni:b}.

Operationally relevant BH features should form at a finite time of a distant observer (Bob).
As trapping of light is  the essence of black holes \cite{curiel},   we formulate the assumption ``a BH exists'' as
a statement  that a trapped region have emerged at some finite time $t_\mS$ of Bob.
The simplest setting to investigate is a spherically-symmetric collapse, where the apparent horizon is unambiguously defined
in all foliations that respect this symmetry \cite{aphor}.
    The analysis of Refs.~\cite{bmmt:18,bmt:18}  produced explicit expressions for the energy-momentum tensor and the metric in the vicinity of  expanding or contracting trapped regions.
 First we briefly summarize the relevant results of Refs.~\cite{bmmt:18,bmt:18} and then explore their implications.

 \section{Geometry in the vicinity of the apparent horizon}    \label{vicinity}
 We    assume validity of semiclassical gravity \cite{pp:09,bmt-1}. That means  we use classical notions (horizons, trajectories, etc.),
 and describe dynamics via the Einstein equations where the standard (or modified) left-hand side is
  equated to the expectation value $T_{\mu\nu}=\6\hat{T}_{\mu\nu}\9_\omega$ of the renormalized stress-energy tensor. The latter represents  both the collapsing matter and the created
excitations of the quantum fields, but we do not assume any specific field state $\omega$.

Boundaries of the trapped region are required to be nonsingular, which is an established property of classical BH horizons.
  We implement this property
 by requiring that the scalars  $\mathrm{T}:=T^\mu_{~\mu}$ and $\mathfrak{T}:=T^{\mu\nu}T_{\mu\nu}$  are finite.
 This is only a necessary condition, and in principle further investigations of the resulting metric are required.
  However, in spherical symmetry these two constraints are sufficient (see Appendix~\ref{apa}  for details).

 Hawking radiation is not assumed. On the contrary,
 the presence of the negative energy density   that is described below is a consequence of the finite formation time of the apparent horizon and its regularity.

    A general spherically symmetric metric   in the Schwarzschild coordinates  is given by
\be
ds^2=-e^{2h(t,r)}f(t,r)dt^2+f(t,r)^{-1}dr^2+r^2d\Omega, \label{sgenm}
\ee
where $r$ is the areal radius. The function $f(t,r)=1-C(t,r)/r$ is coordinate-independent \cite{ms,bardeen:81,aphor}. The Misner-Sharp mass  \cite{bambi,faraoni:b,ms} $C(t,r)$ is invariantly defined  via
\be
1-C/r\defeq \pad_\mu r\pad^\mu r\equiv \nabla_\mu r\nabla^\mu r. \label{defMS}
\ee
This is a gauge-independent equation for a scalar geometrically-defined quantity. On the other hand, the choice of $r$ as one of the coordinates is a partial gauge fixing \cite{aphor}.
 The function $h(t,r)$ plays the role of an integrating factor in transformation to, say, retarded or advanced coordinates.
 In the Schwarzschild spacetime $C=2M=\mathrm{const}$ and $h= 0$.

Trapped regions exist only if the equation $f(t,r)=0$
has a root \cite{krishnam:14}. For any foliation that respects spherical symmetry the areal radius of the apparent horizon $r_\sg$ is found \cite{aphor,hm:66} as a solution of \be
r_\sg=C(r_\sg).
\ee
This root (or, if there are several, the largest one) is the  Schwarzschild   radius $r_\sg(t)$  that identifies the apparent horizon in all such foliations.      For example, in the metric~\eqref{sgenm} an
 outgoing radial null geodesics  has a tangent vector
\be
l^\mu=(1,e^hf,0,0).
\ee
A nonzero coefficient $\kappa\neq 0$  in the parallel transport equation $l^\mu l^\nu_{~;\mu}=\kappa\l^\nu$ is a measure of nonaffinity of the geodesic parametrisation.
Expansion \cite{he:book}  of the congruence of such geodesics is
 \be
 \theta_l=l^\mu_{~;\mu}-\kappa=2e^h (1-C/r)/r.
  \ee
This quantity indeed changes the sign as $r$ crosses $r_\sg$.


  The  assumption of  regularity results    in the generic form of the  energy-momentum tensor  close to the apparent horizon.
   For $x\defeq r-r_\sg(t)\to 0$ its $2\times 2$      block
$a,b=t,r$ is
  \be
T_{ab}=\Xi(t)\begin{pmatrix}
e^{2h} & s e^{ h}/\!f \vspace{1mm}\\
s e^{ h}/\!f & 1/\!f^2
\end{pmatrix},     \quad
 T_{\hat{a}\hat{b}}=   \frac{\Xi(t)}{f}   \begin{pmatrix}
1 & s  \vspace{1mm}\\
s   & 1 \end{pmatrix},
  \label {tneg}
\ee
for some function $\Xi$, and $s=\pm 1$,  and the second expression is written in the orthonormal basis.
This form of $T_{\mu\nu}$ was  obtained in Ref.~\cite{bmmt:18} without using the Einstein equations.
Hence, it will hold in any metric theory, e.g., in $\textsl{f}(R)$ theories \cite{fr:18,fr:t}, in the vicinity of the hypersurfaces
 $f(t,r)=0$.

From now on, we assume that dynamics is described by the standard Einstein equations.
To produce  real solutions with trapped regions at finite time $t$ \cite{bmmt:18} (see Appendix~\ref{apa} for details)
 \be
\Xi(t)=-\Upsilon^2(t)<0,
\ee
must hold, where the function $\Upsilon$ is determined below.  Here $s=\pm 1$ corresponds to $r_\sg'\defeq dr_\sg/dt<0$ and $r_\sg'>0$, respectively. Then the energy-momentum tensor
of Eq.~\eqref{tneg} violates the null energy condition (NEC) \cite{he:book,mmv:17}: $T_{\hat{a}\hat{b}}k^{\hat a}k^{\hat b }<0$
for {a} radial null vector $k^{\hat a}=(1,s,0,0)$.

The metric functions are given  as power series in terms of $x$ as \cite{bmmt:18}
 \be
C= r_\sg(t)-a(t)\sqrt{x}+\frac{1}{3}x\ldots.  \label{c0sin}
\ee
and
\be
h=-\ln\frac{\sqrt{x}}{\xi_0(t)} +\frac{4}{3a}\sqrt{x}+\ldots,\label{h1}
\ee
where $a^2\defeq 16 \pi \Upsilon^2 r_\sg^3$ and the higher-order terms in $x$  depend on the higher-order terms in $T_{\mu\nu}$.
The function  $\xi_0(t)$ is   determined by the  choice of the time variable. 

The function $\Upsilon(t)>0$ is determined by
the rate of change of the Schwarzschild radius,
\be
 r_\sg'/{\xi_0}=\pm4\sqrt{\pi}\,\Upsilon\sqrt{ r_\sg}= \pm a/r_\sg.      \label{lumin}
 \ee

  In the case of a retreating Schwarzschild radius, $r_\sg'(t)<0$, the metric is most conveniently written using the advanced
 null coordinate $v$,
  \be
dt=e^{-h}(dv+ f^{-1}dr). \label{intfu}
\ee
It takes the form of a pure ingoing Vaidya metric,
\be
ds^2=-(1-C_+(v)/r)dv^2+2dvdr +r^2d\Omega, \label{mVv}
 \ee
 where $C_+(v)=C\big(t(v,r),r\big)$ is a decreasing function, $C'_+<0$.
If $r_\sg'(t)>0$ geometry near the apparent horizon is described by a pure  outgoing Vaidya metric
\be
ds^2=-(1-C_-(u)/r)du^2-2dudr +r^2d\Omega, \label{mUu}
 \ee
 where  $C_-'(u)>0$.

Consistency of the Einstein equations allows only two types of the higher-order terms in the components $T_{tt}$, $T^{rr}$ and $T^r_{\,t}$ \cite{bmt:18}.
 In both cases the higher-order terms in both
$h$ and $C$ are monomials of higher half-integer powers of $x$ (Appendix~\ref{apa}).

For a macroscopic black hole ($r_\sg\gg 1$) the evaporation process is quasi-stationary.  The previous analysis should match the steady-state results that are obtained
on a background of an eternal black hole in an asymptotically flat spacetime.
 The steady-state evaporation follows the law $r_\sg'=-\varsigma/r_\sg^2$, where $\varsigma\sim 10^{-3}-10^{-4}$, \cite{bambi,bardeen:81,page:76,bmps:95}. Hence \cite{bmt:18}
\be
\Upsilon\approx\frac{\sqrt{\varsigma}}{2\sqrt{2\pi}r_\sg^2},   \qquad \xi_0\approx2\sqrt{\pi r_\sg^3}\Upsilon=\half a.   \label{page}
\ee

 \section{Physics in the vicinity of the apparent horizon}
 Collapse models   can be   solved only if the matter content and equations of state are known. However, the very fact
 of formation of the apparent horizon allows us to obtain some information about its vicinity.  Consider a radially infalling (not necessarily geodesic) observer Alice that is very close to $r_\sg$.
  Alice's 4-velocity $u_\mathrm{A}^\mu=(\dot T,\dot R,0,0)$
 determines her time axis.  As one of the spacelike directions we take $n^\mA_\mu=e^h(-\dot R,\dot T,0,0)$.
  The energy density and pressure in Alice's frame are always given by $\rho_\mA\defeq T_{\mu\nu}u_\mA^\mu u_\mA^\nu$ and $p_\mA\defeq T_{\mu\nu}n_\mA^\mu n_\mA^\nu$.
 Her 4-velocity is timelike, $u_\mA^\mu u_{\mA\mu}=-1$; hence
  \be
\dot T=\frac{\sqrt{F+\dot R^2}}{e^H F}, \label{t-1}
\ee
where  $F=f\big(T(\tau),R(\tau)\big)$ and $H=h(T,R)$.

This relationships leads to  the comoving values of density and pressure
  close to the retreating $r_\sg$,
  \be
\rho_\mA^<=p_\mA^<=-\frac{\big(\dot R+\sqrt{F+\dot R^2}\,\big)^2}{F^2}\Upsilon^2.
\ee
For $X\defeq R(\tau)-r_\sg\big(T(\tau)\big)\lesssim a^2$  the expansion of $\dot T$ results in  small negative values
\be
\rho_\mA^<=p_\mA^<=-\frac{\Upsilon^2}{4 \dot R^2}+\cO(\sqrt{x}).      \label{denshrin}
\ee
 Using the  metric  of Eq.~\eqref{mVv} that is valid on both sides of a contracting apparent horizon we see that the NEC is violated in some neighbourhood inside the trapped region as well.

 However, in the case of the growing $r_\sg$, when $r_\sg'>0$,
\be
\rho_\mA^>=p_\mA^>=-\frac{\big(\dot R-\sqrt{F+\dot R^2}\,\big)^2}{F^2}\Upsilon^2+\cO (F^{-1}),
\ee
giving a divergent expression
\be
\rho_\mA^>=p_\mA^>=-\frac{2\dot R^2\Upsilon^2}{F^2 }+\cO (F^{-1}),     \label{pdiverp}
\ee
in the vicinity of the apparent horizon, as $F^2\approx a^2 X/r_\sg^2\to 0$.
The flux $\phi\defeq  T_{\mu\nu}u_\mA^{\mu}n_\mA^\nu$ satisfies
\be
\phi_\mA^<=\rho_\mA^<, \qquad \phi_\mA^>=-\rho_\mA^>,   \label{fluxp}
\ee
at the crossing of the retreating and advancing apparent horizons, respectively.

These results show that an expanding trapped region should be accompanied by   firewall---a region of unbounded energy density, pressure and flux---that is perceived by  an infalling inertial observer. Unlike the
  firewall from the eponymous paradox that appears  as a contradiction between four assumptions \cite{firewall,info}, here  it
 is a consequence of regularity of an expanding apparent horizon and its finite formation time.

The comoving values of the matter variables are independent of the function $h(t,r)$.
The divergence follows from the form of the energy-momentum tensor near $r_\sg$ that is given by
   Eq.~\eqref{tneg} and the opposite signs of  $T_{tt}$ and $T_{tr}$. Hence our previous analysis indicates that this divergence occurs in all metric theories.

All the steps that result  in the identification of the metric functions outside the Schwarzschild radius  can be performed in the vicinity of the inner horizon. Then the energy-momentum
 tensor again has the form of Eq.~\eqref{tneg}, but with $\Xi\to +\Theta^2$ for some $\Theta(t)$. The solution of the Einstein equations has a similar form, and
 for the inner horizon propagating towards
  the center, $r_\mathrm{in}'<0$,
 we find that $\pad_t C>0$ (and divergent, as $r$ approaches $r_\mathrm{in}$ from below). Hence $0<T_t^r=+\Theta^2$. For a comoving observer
 Charlie that is overtaken by the inner horizon the local density, pressure and    flux are
 \be
 \rho_\mathrm{C}=p_\mathrm{C}=\phi_\mathrm{C}=+  \frac{\Theta^2}{4 \dot r_\mathrm{C}^2}.     \label{inrho}
 \ee

 \subsection{Horizon crossing by test particles}
A massive test particle will cross the apparent horizon when the gap  \cite{bmt-1,kmy:13}
 \be
 X(\tau)\defeq R(\tau)-r_\sg\big(T(\tau)\big),   \label{gap1}
 \ee
 becomes zero.
The crossing is prevented if for some $X>0$  (and $r_\sg>0$)
\be
\dot X=\dot R-r_\sg'\dot T>0.
\ee
 An analogous expression holds for the outgoing Vaidya metric \cite{bmt-1,kmy:13}, but not for the   ingoing Vaidya metric of \eqref{mVv} \cite{uac:18,bmt:18}.

 In the vicinity of the apparent horizon $x\ll r_\sg$ and
\be
\dot T\approx -\dot R e^{-H}/F.       \label{timeap}
\ee
  Using Eqs.~\eqref{c0sin} and \eqref{h1}   we find that
    \be
    \dot X=-\frac{(\dot R^2-4\pi r_\sg^2\Upsilon^2)}{2|\dot R|\sqrt{\pi}r_\sg^{3/2}\Upsilon}\sqrt{X}+\cO(X),
    \ee
    and see that if a test particle is in the vicinity of the apparent horizon, $X\ll a^2$,  it will cross the horizon unless $|\dot R| <2\sqrt{\pi}r_\sg\Upsilon\sim \sqrt{\kappa}\sim 0.01$.
 (For comparison, a free-falling
particle starting at rest from infinity crosses the event horizon of a classical black hole with $\dot R=-3/4$).    This difficulty of crossing the horizon for slow-moving test particles is consistent with the
results of Ref.~\cite{bbgj:16}.

Using the leading higher-order terms in the metric functions (Appendix~\ref{apa}) allows to obtain
 terms of the order of $X$ and $X^{3/2}$ in  the expansion of $\dot X$.  Their evaluation under
assumption of the quasi-stationary  evaporation does not lead to qualitatively different conclusions.

 The same analysis applies to massless test particles.  In this case the trajectory is most conveniently parameterized by $\lambda\defeq -R$  \cite{mnt:18}, and one evaluates the derivative $dX/d\lambda$.  Then
 $\dot R\to dR/d\lambda\equiv -1$, and the apparent horizon is always crossed.

\subsection{Horizon dynamics}
A general spherically-symmetric metric  in comoving coordinates (here comoving means that
fictitious freely-falling observers remain at fixed values of  the spatial coordinates  $\chi$, $\theta$, $\phi$) is given by \be
ds^2=-e^{2\lambda}d\bar t\,^2+e^{2\psi}d\chi^2 +r^2d\Omega^2, \label{comet}
\ee
where the areal radius $r(\bar t,\chi)$  and the functions $\lambda(\bar t,\chi)$
and $\psi(\bar t,\chi)$ are to be determined.      For an observer at $\chi=\mathrm{const}$,
 the proper time is given by $d\tau=e^{\lambda }d\bar t$, and the outward-pointing spacelike normal is $n_\mu=(0,e^\psi,0,0)$. Then the comoving energy density, pressure,   and flux are
\be
\rho=-T^{\bar t}_{~\bar t}, \qquad p= T^\chi_{~\chi}, \qquad  \phi= T^\chi_{~\bar{t}} e^{\psi-\lambda}.     \label{comovT}
\ee

 The Misner-Sharp mass $\eC(\bar t,\chi)$  [defined via Eq.~\eqref{defMS}] simplifies the Einstein equations \cite{bambi,m-r}. In the metric \eqref{comet} it is
\be
\eC(\bar t,\chi)=r\big(1-e^{-2\psi}(\pad_\chi r)^2+e^{-2\lambda}(\pad_{\bar t}r)^2\big),     \label{massC}
\ee
and the three   relevant  Einstein equations  are
\begin{align}
&-\frac{\pad_\chi \eC}{r^2\pad_\chi r}+
\frac{2\hspace{1pt}\pad_{\bar t}\,r e^{-2\lambda}}{r\pad_\chi r}\big(\pad_{\bar t}\pad_\chi r -\pad_{\bar t}\hspace{1pt} r\pad_\chi\lambda- \pad_{\bar t}\hspace{1pt}\psi \pad_\chi r\big)
 =-8\pi \rho,    \label{poc}    \\
& -\frac{\pad_{\bar t} \eC}{r^2 \pad_{\bar t} r}-
\frac{2\hspace{1pt}\pad_\chi r e^{-2\psi}}{r\pad_{\bar t}\,r }\big(\pad_{\bar t}\pad_\chi r -\pad_{\bar t}\hspace{1pt} r\pad_\chi\lambda- \pad_{\bar t}\hspace{1pt}\psi \pad_\chi r\big)
 =8\pi p,  \label{doc}\\
& -\frac{2\hspace{1pt} }{r }\big(\pad_{\bar t}\pad_\chi r -\pad_{\bar t}\hspace{1pt} r\pad_\chi\lambda- \pad_{\bar t}\hspace{1pt}\psi \pad_\chi r\big)
=8\pi\hspace{-1pt}\phi\hspace{1pt} e^{\lambda+\psi}.     \label{greek}
\end{align}

In contrast, the simplest models of gravitational collapse  describe matter as a single perfect fluid with   a comoving energy-momentum tensor
\be
T^\mu_{~\nu}=\mathrm{diagonal}(-\rho,p,p,p).
\ee
The absence of the  flux term, $\phi\equiv 0$, leads via Eqs.~\eqref{greek} and \eqref{poc} to a compact expression for the  mass,
     \be
     \eC(\bar t,\chi)= 8\pi\int_0^\chi \rho r^2r'd\chi\equiv C\big(t(\bar t,\chi),r(\bar t,\chi)\big), \label{masse}
     \ee
where the last identity follows from the definition \eqref{defMS} evaluated in $(t,r)$ coordinates with the metric of Eq.~\eqref{sgenm}.
However, 
at the
apparent horizon the flux is as
important as pressure. Models that involve several nonideal fluids \cite{dl:68,rhyd} should be used to describe the BH formation at finite Bob's time.

Violations of the NEC are bounded by quantum energy inequalities (QEIs) \cite{few:17}. For spacetimes of small curvature explicit expressions that bound time-averaged energy density for a geodesic observer
  were derived in Ref.~\cite{eleni}.
   For any Hadamard state $\omega$ and a sampling
function $\mathfrak{f}(\tau)$ of compact support,
 negativity of the expectation  value of the energy density  $\rho=\6 \hat{T}_{\mu\nu}\9_\omega u^\mu u^\nu$ as seen by a geodesic observer that moves on a trajectory $\gamma(\tau)$ is bounded by
\be
 \int_\gamma\! \mathfrak{f}^2(\tau)\rho d\tau \geqslant - B(R,\mathfrak{f},\gamma), \label{qei}
\ee
where $B>0$ is a bounded function that depends on the trajectory, the Ricci scalar and the sampling function \cite{eleni}.

Consider a growing apparent horizon, $r_\sg'>0$.
 For a macroscopic BH the curvature at the apparent
 horizon is low and its radius does not appreciably change while Alice moves in the domain of validity of Eq.~\eqref{pdiverp}. Then $\dot X \approx \dot R$, and given Alice's
 trajectory we can choose  $\mathfrak{f}\approx 1$ at the horizon crossing and $\mathfrak{f}\to 0$  within the NEC-violating domain. As the trajectory
 passes through  $X_0+r_\sg\to r_\sg$ the lhs of Eq.~\eqref{qei} behaves as 
 \be
 \int_\gamma\! \mathfrak{f}^2\rho_\mA d\tau\approx -\int_\gamma\frac{\dot R^2d\tau}{8\pi r_\sg\, X}\approx \int_\gamma\frac{|\dot R|dX}{8\pi r_\sg\, X}\propto \log X_0\to-\infty,
 \ee
where we used $\dot R\sim\mathrm{const}$. The rhs  of Eq.~\eqref{qei} remains finite, and thus the QEI is violated.  This violation indicates  the apparent horizon cannot grow.

\begin{figure}[htbp]
\includegraphics[width=0.28\textwidth]{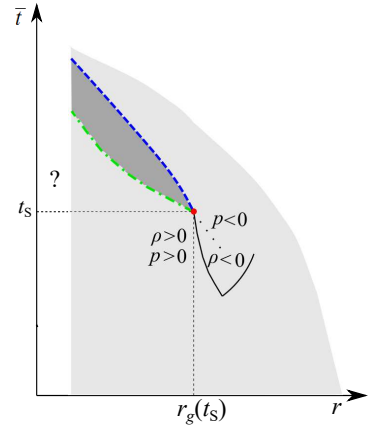}
\caption{Schematic structure of the early stages of the evolution of a trapped region (dark gray) if it forms at finite time $t_\mS$.
Possible structures in the white patch near the time axis are not constrained by our considerations. The blue dashed line represents the
 apparent horizon, the green dot-dashed line represents the inner apparent horizon.
 The first marginally trapped surface at $r=r_\sg(t_\mS)$ is marked as a red dot. Part of the region of negative density is outlined by a thin
 black line.  Part of the boundary of the region of negative pressure is marked
 by the dotted black line. The shape and end points of the last two lines are not constrained by our considerations.}
\label{schema1}
\end{figure}

 The comoving density and flux are finite on the approach to the receding apparent horizon and the comoving metric remain regular. We can write Eq.~\eqref{doc} as
\be
 -\frac{e^{-\lambda}\pad_{\bar t} \eC}{r^2 \dot r_\chi}+ 8\pi\frac{\pad_\chi r}{\dot r_\chi}\phi\hspace{1pt} e^{-\psi}      =8\pi p,   \label{rewritedtC}
 \ee
  where $\dot r_\chi=\pad_{\bar t}r(\bar t,\chi)d\bar t/d\tau<0$,
   and the subscript $\chi$ indicates that the areal radius corresponds to a fixed comoving coordinate.  From Eq.~\eqref{c0sin} it follows that   both $\pad_t C$ and $\pad_r C$ diverge as  $1/\sqrt{r-r_\sg}\,$.
 Using
  \begin{align}
\pad_{\bar t} \eC &= \left.\pad_t C(t,r)\frac{\pad t}{\pad \bar t}\right|_\chi+ \left.\pad_r C(t,r)\frac{\pad r}{\pad \bar t}\right|_\chi       \nonumber \\
&=e^{\lambda} \big( \pad_t C(t,r)\dot t_\chi + \pad_r C(t,r)\dot r_\chi\big),            \label{ccgrow}
\end{align}
we find that $ \pad_{\bar t} \eC$ 
is finite only if $r'_\sg\, \dot t_\chi=\dot r_\chi$ as $r\to r_\sg$ (see Appendix~\ref{apb} for details).  However, Eqs.~\eqref{timeap} and \eqref{lumin}
imply precisely this relation, and thus no infinities are necessary to satisfy the Einstein equations  \eqref{poc}--\eqref{greek}.

\section{Conclusions}
 Trapped regions are physically relevant only if their formation time is finite. Hence the only assumption
 we have made is the regularity of their boundary.  We find that
the NEC is violated in the vicinity  of the apparent horizon and is satisfied in the vicinity of the inner apparent horizon.
The form of the energy-momentum tensor that is given by  Eq.~\eqref{tneg} is the same in all metric theories of gravity,
 not    only in GR. We expect that the NEC violation is also a necessary condition for the finite-time (according to Bob) formation of  trapped regions  in $\textsl{f}(R)$ theories,
 and we will investigate them in a future work.

Flux cannot be neglected in the vicinity of the apparent horizon. Hence the collapse of a single ideal fluid (even allowing for violation of the NEC)  cannot lead to
 formation of a black hole in finite time of a distant observer.
 In the classical homogeneous collapse the first marginally trapped surface appears at the boundary of the collapsing body. 
 However,  the Misner-Sharp
mass  $\eC>0$, while the energy density is negative on both sides of the apparent horizon,
 no  system   with a uniform density can form it.

Expanding the apparent horizon precipitates a firewall. Its divergent density, pressure and flux  do not lead to singularities,
 but violate the QEI. Hence either trapped regions cannot grow  or the semicalssical analysis is inapplicable in their vicinity  even if the curvature is small, as argued in Ref.~\cite{qua-mod}. It has
 a simple intuitive explanation: growth of $r_\sg$ means growth of the BH mass. However, only the NEC-violating matter with negative energy density can cross the horizon, contributing to the mass decrease.

Infalling massive test particles may and massless test particles will cross the apparent horizon. However, the proverbial dropping of the Encyclopedia Britannica into a black hole that is
followed by the alleged loss of information is impossible. A mandatory violation of the NEC in some vicinity of the apparent horizon is incompatible with preservation
of the normal character of the perturbing matter.  Hence we have to investigate how perturbations by normal matter evolve and what happens to the perturbing material.

 Propagating the limits of $\rho$, $p$ and $\phi$ back to $t_\mS$ show that the first marginally trapped surface  is a surface of discontinuity of the properties of collapsing matter,
 and a rather complicated   diagram (Fig.~1) emerges. When the trapped region formes the density in the central region of the
 collapsing body is still positive, $\rho>0$ for $r\leq r_\mathrm{in}$.
Causality and/or continuity arguments   in the vicinity of $r_\sg(t_\mS)$ indicate that  energy density becomes negative in some region close to $r_\sg(t_\mS)$
before formation of the horizons. If for some $t_0<t_\mS$
it  coincides with  the region of $p<0$, as well as       the possibility of  discontinuity and shock waves inside the trapped region
  will be investigated.

 \acknowledgments
 Useful discussions with Pisin Chen, Eleni Kontou,  Robert Mann, Sebastian Murk and Mark Wardle are gratefully acknowledged.

\appendix
\section{Solutions of the spherically-symmetric Einstein equations near an apparent horizon} \label{apa}

In spherical symmetry the trace and the square of the energy-momentum tensor are
 \begin{align}
    \mathrm{T}=&-e^{-2h }T_{tt}/f  +T^{rr}/f +2T^\theta_{~\theta}, \label{fin1}\\
  \mathfrak{T}=&-2\left(\frac{e^{-h} T_t^{\,r}}{f }\right)^2+\left(\frac{e^{-2h }T_{tt}}{f }\right)^2
    +\left(\frac{T^{rr}}{f }\right)^2 +2\big(T^\theta_{~\theta}\big)^2.       \label{fin2}
    \end{align}
Assuming that $T^\theta_{~\theta}$ is finite (this can be proven in case of the standard GR), we obtain Eq.~\eqref{tneg} as a generic case \cite{bmmt:18}.

The  Einstein equations that determine the functions $h$ and $C$ in the Schwarzschild coordinates  are
\begin{align}
G_{tt}=&\frac{e^{2h}(r-C)\pad_r C}{r^3}=8\pi T_{tt}, \label{gtt}\\
G_t^{\,r}=&\frac{\pad_t C}{r^2}=8\pi T_t^{\,r}, \label{gtr}\\
G^{rr}=&\frac{(r-C)(-\pad_r C+2(r-C)\pad_r h)}{r^3}=8\pi T^{rr}. \label{grr}
\end{align}

The requirement that the  scalars   $\mathrm{T}$ and $\mathfrak{T}$ are finite
 leads to the form of the energy-momentum tensor that is given by Eq.~\eqref{lumin}. The negative sign that results in  violation of the NEC is necessary for having
 real solutions of the Einstein equations. This result should be compared with the conclusions of Sec 9.2 of Ref.~\cite{he:book} that in general asymptotically
flat spacetimes with an asymptotically predictable future  the trapped surface
cannot be visible from the future null infinity unless the weak energy condition is violated \cite{he:book,fn:book}.
 Here  we have considered only a spherically-symmetric setting, but without making assumptions about asymptotic structure of the spacetime \cite{bmmt:18}.

Working in $(u,r)$  or $(v,r)$ coordinates provides the easiest way to establish that other curvature scalars are finite.
 In a general four-dimensional spacetime there are 14 algebraically
 independent scalars that can be constructed from the Riemann tensor \cite{exact:b,zm:97}. A   convenient system of polynomial invariants consists of the Ricci scalar and further 15 invariants
 \cite{cm:91}. A direct calculation  \cite{ccgrg} shows that  for the metrics \eqref{mVv} and \eqref{mUu} all the invariants are identically zero, except for the two finite invariants
 that are constructed using the complex conjugate  of the self-dual Weyl tensor \cite{cm:91},
 \be
  \bar{C}_{\kappa\lambda\mu\nu} :=\half( C_{\kappa\lambda\mu\nu} +i{}^*\!C_{\kappa\lambda\mu\nu}),
  \ee
 the invariants being
 \be
 W_1:=\tfrac{1}{4}\bar{C}_{\kappa\lambda\mu\nu} \bar{C}^{\kappa\lambda\mu\nu},
 \qquad   W_2:=\tfrac{1}{4}\bar{C}_{\kappa\lambda\mu\nu} \bar{C}^{\mu\nu}_{~~\rho\sigma} \bar{C}^{\rho\sigma\kappa\lambda}.
  \ee
  It is also easy to see that in this metric all the components of the Riemann tensor in the Vaidya coordinates (Eqs.~\eqref{mVv} and \eqref{mUu}) are finite at $r=r_\sg$.

The higher-order terms in metric and energy momentum are of one of the two possible types \cite{bmt:18}. The series expansion can be either regular,
\be
 T_{\hat{a}\hat{b}}{f}        =-\Upsilon^2 +\sum_{n>1}\alpha^{(ab)}_n x^n, \ee
or regular singular,
 \be
 T_{\hat{a}\hat{b}}{f}        =-\Upsilon^2 +\sum_{n>1}\alpha^{(ab)}_n x^{n-1/2}.
 \ee

 In both cases  the expansion follows the same pattern,
 \be
C(t,r_\sg+x)=r_\sg-a\sqrt{x}+\tfrac{1}{3}x+c x^{3/2}+g x^2+\ldots
\ee
and
\be
h(t,r_\sg+x)=-\ln\frac{\sqrt{x}}{\xi_0}+k_2 \sqrt{x}+k_3 x+k_4 x^{3/2}+\ldots,
\ee
 For a regular correction to $T_{\mu\nu}$ (we set $\alpha^{(tt)}_1=\alpha_1$,  $\alpha^{(tr)}_1=\beta_1$,   $\alpha^{(rr)}_1=\gamma_1$),
 the terms of the metric functions that depend only on first-order corrections are \begin{align}
a&=4 \sqrt{\pi } r_\sg^{3/2}\Upsilon, \\
 c&= \frac{ \left(36 \pi
\alpha_1 r_\sg^3-108 \pi  r_\sg^2 \Upsilon^2-1\right)}{36 \sqrt{\pi } r_\sg^{3/2}
\Upsilon},\\
 g&=\frac{1}{540} \left(-\frac{36 \alpha_1}{\Upsilon^2}+\frac{1}{\pi  r_\sg^3
\Upsilon^2}+\frac{108}{r_\sg}\right).
\end{align}
and
  \begin{align}
k_2 &= \frac{4}{3 a},\\
k_3 &= - \frac{3}{2r_\sg}-\frac{c}{a}+\frac{24 \pi  \alpha_1 r_\sg^3+24 \pi  \gamma_1 r_\sg^3-4  }{6 a^2},\\
k_4 &= \frac{2 \left( 27 a^2 g - 54 a c - 16 \right)}{81a^3} \nonumber \\
& \quad + \frac{2 \left( -54 a^2 + 144 \pi \alpha_1 r_\sg^4 + 144 \pi \gamma_1 r_\sg^4 \right)}{81a^3 r_\sg},
\end{align}
where the functions $\xi_0(t)$  and $\Upsilon(t)$ are given by Eq.~\eqref{page}.  Using Eq.~\eqref{gtr}  and the conservation law $\nabla_\mu T^\mu_{~\nu}=0$ for $\nu=0,1$ allows to obtain the
recursive relations for  the higher-order coefficients $\alpha^{(ab)}_n$ \cite{bmt:18}.

\section{Details of Eq.~\eqref{ccgrow}}     \label{apb}

Consider now the receding apparent horizon, $r_\sg'<0$.  The invariants $\mathrm{T}$ and  $\mathfrak{T}$ are  finite. 
In general it does not imply that the metric functions are regular as $r\to r_\sg$: the functions $h(t,r)$ and $\pad_t C(t,r)$ diverge.

      However, the Einstein equations imply \begin{align}
\pad_{\bar t} \eC &= \left.\pad_t C(t,r)\frac{\pad t}{\pad \bar t}\right|_\chi+ \left.\pad_r C(t,r)\frac{\pad r}{\pad \bar t}\right|_\chi       \nonumber \\
&=e^{\lambda} \big( \pad_t C(t,r)\dot t_\chi + \pad_r C(t,r)\dot r_\chi\big),            \label{ccgrow-a}
\end{align}
   Here, we present in detail the analysis of Eq.~\eqref{ccgrow}. The two partial derivatives of $C(t,r)$ are
\be
\pad_t C=\frac{2}{3} r_\sg'-a'\sqrt{r-r_\sg}+\frac{a r_\sg'}{2 \sqrt{r-r_\sg}},
\ee
 and
 \be
 \pad_r C=-\frac{a r_\sg'}{2 \sqrt{r-r_\sg}} +\frac{1}{3},
 \ee
 where we omitted   terms that approach zero as $r\to r_\sg$.
For $\lambda>-\infty$ and $r_\sg'<0$ Eq.~(L-33) implies that the derivative $        \pad_{\bar t} \eC $ at $r_\sg$,
is finite,
i. e.
\be
\left|\lim_{r\to r_\sg}\left( \frac{a r_\sg'\,\dot t_\chi}{2 \sqrt{r-r_\sg}} -     \frac{a \dot r_\chi}{2 \sqrt{r-r_\sg}}  \right)\right|<\infty,
\ee
only if
\be
\lim_{r\to r_\sg}(r_\sg'\,\dot t_\chi-  \dot r_\chi)=0,
\ee
 i.e.
 \be
 r_\sg'\,\dot t_\chi\to \dot r_\chi, \label{cond2}
 \ee
and the difference goes to zero faster than $\sqrt{x}$. Since the trajectory of a comoving particle is timelike, expansion of Eq.~(L-12) results in
\be
\dot t_\chi=- \frac{\dot r_\chi}{4\sqrt{\pi r_\sg} \,\xi_0\Upsilon}  +\cO(\sqrt{r_\chi-r_\sg}),
\ee
which  for a retreating apparent horizon   implies Eq.~\eqref{cond2} via the consistency condition Eq.~(L-7).

 Using this relationship we find
  \be
\lim_{r\to r_\sg}   e^{-\lambda}  \pad_{\bar t} \eC=\frac{2}{3}r_\sg'\dot t_\chi+\frac{1}{3}  \dot r_\chi   =r_\sg'\,\dot t_\chi<0,
\ee
and the limiting form of Eq~(32) becomes
\be
-\frac{1}{r_\sg^2}+8\pi \frac{\pad_\chi r}{\dot r_\chi}\phi e^{-\psi}\approx 8\pi p, \label{simsim}
\ee
in the vicinity of the apparent horizon both for $r_\sg'<0$ and $r_\sg'>0$.

In the former case $\phi\approx p<0$. Using the approximation  $r_\sg'=-\kappa/ r_\sg^2$ to express the matter variable Eq.~\eqref{simsim} becomes
\be
 -\frac{1}{r_\sg^2}+\frac{\pad_\chi r}{|\dot r^3_\chi|}\frac{\kappa e^{-\psi}}{r_\sg^4}\approx-\frac{\kappa}{r_\sg^4},
\ee
 and the equation is satisfied if the function $\psi$ reaches a large negative (but finite) value.       In the latter case $\phi\approx -p\propto 1/x>0$,
 and Eq~\eqref{simsim} becomes
 \be
  \frac{\pad_\chi r}{\dot r_\chi} e^{-\psi}\approx -1.
 \ee
 This relation indicates that unless    the so-called shell crossing singularity \cite{bambi} occurs, $\pad_\chi r=0$, the function $\psi$ should satisfy $\psi>-\infty$.

\end{document}